\documentclass[twoside,11pt]{article}

\usepackage{obs_study_style}
\usepackage[colorlinks,allcolors=black]{hyperref}
\usepackage{natbib}  
\heading{}{}{}{}{}{Xuming He and Jingshen Wang}

\ShortHeadings{Comments on ``Two Cultures"}{He and Wang}
\firstpageno{1}

\begin{document}

\title{Comments on ``Two Cultures": What have changed over 20 years?}

\author{\name Xuming He \email xmhe@umich.edu \\
       \addr Department of Statistics\\
       University of Michigan\\
       273 West Hall, Ann Arbor, MI 48109, USA
       \AND
       \name Jingshen Wang \email jingshenwang@berkeley.edu \\
       \addr Division of Biostatistics\\
       University of California, Berkeley\\
   		5408 Berkeley Way West, CA 94710, USA}

\maketitle

\begin{abstract}%   <- trailing '%' for backward compatibility of .sty file
Twenty years ago \cite{breiman2001statistical} called to our attention a significant cultural division in modeling and data analysis between the stochastic data models and the algorithmic models. Out of his deep concern that the statistical community was so deeply and ``almost exclusively" committed to the former, Breiman warned that we were losing our abilities to solve many real world problems. Breiman was not the first, and certainly not the only statistician, to sound the alarm; we may refer to none other than John Tukey who wrote almost 60 years ago ``data analysis is intrinsically an empirical science". However, the bluntness and timeliness of Breiman's article made it uniquely influential. It prepared us for the data science era and encouraged a new generation of statisticians to embrace a more broadly defined discipline.    

Some might argue that ``The cultural division between these two statistical learning frameworks has
been growing at a steady pace in recent years", to quote   \cite{wang2020breiman}. In this commentary, we focus on some of the positive changes over the past 20 years and offer an optimistic outlook for our profession.\\

{\it Keywords:} Algorithm; Data models; Machine learning, Prediction.
\end{abstract}

\section{Diverse evaluation metrics are taking root in statistics}

Classical statistical theory relies mostly on probability or stochastic models and places much emphasis on the accuracy of parameter estimation (e.g., bias, variance, mean squared error, coverage probability), and the validity and the power of hypothesis testing on certain parameters of interest. However, we also routinely use  predictive/forecasting accuracy, classification accuracy, individual/subgroup treatment effects, receiver operating characteristic curves and so on in statistical assessments, depending on which is more relevant to the problem being addressed.

The importance of prediction is gaining recognition. We found the use of ``prediction error" in 16 articles (including comments and book reviews) published in {\it Journal of the American Statistical Association } over a 3-year period of 1998-2000, and the number rose more than quintupled to 86 over the most recent 3-year period (2018-2020). To put this in perspective, the same search engine showed an increase from 79 articles to 138 (just shy of doubling) over the same time period for ``logistic regression". 

Prediction is certainly not the only thing that matters. We do not think it is fair or even possible to quantify how much we do in statistics should be about prediction accuracy. Instead, we ask whether more needs to be done to promote the use of modern statistical and machine learning methods in real world applications. We share our experience in two areas of applications.

Sport concussion concerns many but accurate diagnosis and good understanding of recovery and of longer term impact remain challenging. Because concussion is typically a result of biomechanical forces transmitted to the cerebral tissues from impacts to the head or torso, clinical researchers have asked whether biomechanical analyses of impacts can help in developing a diagnostic model and improving protective equipment for the athletes. Earlier work using tree-based methods \citep{broglio2010biomechanical} to analyze data collected from the Head Impact Telemetry System helped quantify importance of the biomecehnical measurements and the relevant thresholds  for concussive injuries sustained during high school football. Over the years we also learned that accuracy of concussion diagnosis should not rely on those measurements alone. Furthermore, machine learning methods including the deep neutral network model was introduced to clinical researchers for predicting time-to-recovery in pediatric concussion patients; see \cite{walker2019machine}. %Walker et al. (2019, $doi.org/10.1542/peds.144.2\_MeetingAbstract.198$). 
Genetic Fuzzy Trees (GFT), a modern artificial intelligence tool, was employed in \cite{fleck2020predicting} to  predict post-concussion symptom recovery in adolescents. When compared with common linear classification methods, the GFT was shown to have significantly better overall accuracy than most of its competitors based on cross-validated measures of sensitivity and specificity. In this particular study the overall accuracy based on Discriminant Function Analysis (DFA) and Random Forest was 54.4\% and 57.1\%, respectively, but increased to 62.3\% when the GFT was used. Studies of this nature in concussion research are still limited, but our engagement in the Michigan Concussion Center has convinced us that our colleagues often turn to us as statisticians for help with both exploratory and confirmatory analysis, and few of them  consider us as experts of "one culture". They talk to us about predictive accuracy as well as significance tests under interpretable models.

Cardiovascular disease (CVD) is a major cause of death worldwide. Multiple risk factors (i.e, age, gender, smoking status, body mass index, systolic blood pressure, history of diabetes, and reception of treatments for hypertension) are associated with CVD. Appropriately accounting for these risk factors is known to improve accuracy of CVD risk prediction and hence extend lives. Conventional methods including the Cox proportional hazard models are still being frequently employed to estimate the CVD risk based on these familiar risk factors \citep[see][for example]{pylypchuk2018cardiovascular, kaptoge2019world}. In fact, many current guidelines often recommend calculating the CVD risk score based on a limited number of predictors: the 2010 American College of Cardiology/American Heart Association (ACC/AHA) guideline recommended the Framingham Risk Score \citep{greenland20102010}, the 2016 United Kingdom National Institute for Health and Care Excellence (NICE) guidelines used the QRISK3 score \citep{hippisley2017development}, and the 2016 European Society of Cardiology (ESC) guidelines included the SCORE model \citep{massimo2016european}. 

Because of restrictive modelling assumptions adopted in traditional CVD risk score models, these approaches generally exhibit modest predictive performance. In recent years, modern machine learning tools have begun to be employed for predicting the CVD risk in the hope of agnostically discovering novel risk predictors and learning the complex interactions between them \citep[see][for a comprehensive review]{krittanawong2020machine}. For example, ``AutoPrognosis", an  automated clinical prognostic modeling tool, was able to correctly predict 368 more cases out of 4,801 CVD cases than a traditional approach in analyzing UK Biobank data; see \cite{alaa2019cardiovascular}. It was shown in \cite{sung2019development} that a deep learning prediction model provided some promising results in predicting the CVD risk. There, the deep learning model had time-dependent ``Area under the ROC Curve" (AUC) of as high as 0.94 for females and 0.96 for males, while the classical Cox regression model showed time-dependent AUC of up to 0.79 for females and 0.75 for males. In both studies cited above, the contribution of each variable to the prediction was used to rank the possible risk factors, but a variable that contributes more to prediction is not necessarily causal. Quantification of a risk factor via a causality study in the machine learning framework remains underdeveloped. Nevertheless, the capacity of machine learning algorithms to model highly interactive complex data structures in CVD has gained momentum. By Google Scholar counts, we found that around 10\% of the publications about CVD discussed machine learning methods in 2010, and the share has risen to about one third by now.

\section{Data models and algorithmic models are no longer easily distinguishable}

Breiman made a convincing case that stochastic data models used in classical statistics are not the best, or even appropriate, models for many real world problems.  Algorithmic models often offer a much more flexible framework for data analysis. We would like to emphasize that data models in statistics are often used as working models to generate a procedure or algorithm. They do not need to be correct or even accurate descriptions of the true nature, either for prediction or for inference. In this sense, data models and algorithmic models are often two starting points that could end up with friendly hand-shaking.

There are many examples where mis-specification of the data models are taken into account in inference. Generalized estimating equations (GEE), originally designed to account for a possible unknown correlation between outcomes \citep{zeger1986longitudinal}, %(Liang and  Zeger, 1986, doi:10.1093/biomet/73.1.13), 
relies on specification of only the moments in a data model. By using a robust variance-covariance estimates, statistical inference on the model parameters remain asymptotically valid without stringent assumptions assumed on part of the models.
Another example is the posterior inference in a Bayesian framework. \cite{yang2016posterior} %(2016,  doi.org/10.1111/insr.12114) 
showed that by adopting a convenient working likelihood for quantile regression, Markov Chain Monte Carlo algorithms can then be used to draw from the posterior distribution to form a basis for asymptotically valid inference on the quantile regression parameter; here the assumed data model is really used as an input to the Bayesian computational framework, and the entire inferential process can be viewed as an algorithm.  

Taking it a little further, we have mixture of experts \citep{jacobs1991adaptive} %(Jacobs et al. 1991, doi.org/10.1162/neco.1991.3.1.79) 
that starts with Gaussian mixture models. The Gaussian components may allow us to derive and study certain properties of the model but we can take this supervised learning procedure as algorithmic.

In addition, current research in analyzing the CVD risk often finds data models and algorithmic models complimentary in identifying important risk factors. When predicting the risk of CVD by using deep learning models, \cite{sung2019development} ranked the influence of the input variables by using a Layer-wise Relevance Propagation (LRP), one of many explainable artificial intelligence techniques used in neural networks. They found that many risk factors (including age, gender, blood pressure, smoking, exercise) considered to be important in classical clinical studies were also shown to be important by deep learning models.  Deep learning has its own limitation in providing the effect size of those risk factors due to hidden layers used in LRP.  Nevertheless, \cite{alaa2019cardiovascular} found that the algorithmic models helped uncover novel risk factors for CVDs that have not been considered in traditional prospective studies when analyzing the UK Biobank data. For women, AutoPrognosis identified ``ankle spacing width" as an important risk factor. Clinically, this may be linked to symptoms of poor circulation, such as swollen legs, which could be predictive of future CVD events. For patients with history of diabetes, AutoPrognosis has not only maintained high predictive accuracy but also revealed diabetes-specific risk factors (such as microalbuminuria) that were not previously captured by the traditional data models.

We certainly do not intend to claim that the difference between the data models and the algorithmic models no longer exists. It remains obvious that both types of models can be mis-used. In fact, dangers have persisted in many statistical analyses that rely too much on blind trusts of the data models, which has led to the reproducibility and replicability crisis as some have called it. There are also many instances where blind use of algorithmic models and machine learning methods have failed us. The Google Flu Trends \citep{butler2013google}, IBM’s Watson supercomputer cancer treatment recommendations \citep{ross2018ibm}, and Apple Card Algorithm \citep{vigdor2019apple, pena2020bias} are just some of well-publicized examples. We do not think that these problems argue against either type of models; rather their co-existence and further promotion of their better use make statistics a better playground. This leads us to the next point.

\section{Statistical learning needs domain expertise}

Data analysis is not just about data. Instead, any modeling attempt needs to incorporate domain knowledge and expertise wherever possible. We think this is a bigger issue than which modeling approach one takes.

Take statistical downscaling as an example. Statistical downscaling aims to localize global or regional climate model projections to assess the potential impact of climate changes based on data-driven models. It requires quantifying a relationship between climate model outputs ($X$) and local observations ($Y$) from the past. One would naturally attempt to regress $Y$ on $X$ to find a relationship that can be used for future projections of $Y$ based on the climate model output. However, the projections from typical climate models are about ``climate", not daily/weekly/monthly measurements that pair with the local observations in real time. Without the domain knowledge that we had to learn from the atmospheric scientists, we would run the risk of making poor choices of statistical modeling tools, whether it is a simple linear regression or a more complicated machine learning method. In fact, we need to turn to regression-type modeling with asynchronous measurements as demonstrated in \cite{he2012bivariate} and \cite{stoner2013asynchronous}. 
%He et al. (2012, doi.org/10.1007/s13253-012-0098-6) and Stoner et al. (2013,  doi.org/10.1002/joc.3603).

In computational medicine, it is no surprise that machine learning algorithms without incorporating  domain knowledge can make mistakes that would appear trivial to domain experts. In a recent study, CheXNet (Radiologist-Level Pneumonia Detection on Chest X-Rays with Deep Learning), \cite{rajpurkar2017chexnet} observed that a convolutional neural network (CNN) outperformed radiologists in overall diagnostic accuracy. A subsequent study \citep{zech2018variable} revealed that the CNN built in CheXNet suffers from high generalization bias. Part of the reasons is that the CNN chooses some hospital site-specific variables over patient underlying pathology as predictive variables.  Such a choice was made based on a strong correlation displayed in CheXNet between the pneumonia prevalence and those site-specific variables. In addition to the site-specific confounding variables that threaten the generalizability of the CNN, there can be other factors related to medical management that undermines the clinical applicability of statistical learning methods. \cite{oakden2020exploring} pointed out since chest tubes are frequently used to treat pneumothorax, a CNN without context-oriented modelling may not be able to diagnose patients with pneumothorax since they lacked a chest tube. To make  statistical learning methods more trustworthy in medical research, adapting to domain knowledge is an important step towards building stronger and more robust models. Whether one calls this ``interpretable  machine learning" or ``explainable AI", we need to be embedded in the relevant domain of science, business, or medicine.

Data-driven models and prediction have proven to be useful but not without peril. Things could go wrong easily if data are taken out of context. %The initial success and the spectacular failure of Google Flu Trends  served multiple reminders to all. 
To conclude, we reiterate some of the points made earlier.
First, data alone, no matter how big, may not tell the whole story. Any prediction trained on the existing data without sanity checks based on domain knowledge or even common sense could turn into a fiasco. Second, a good model needs to incorporate the dynamic nature of the world, where non-stationarity and heterogeneity of various forms cannot be ignored. Generalization errors that do not look beyond the past data are not always trusty measures. Third, we need to understand the limitations of what we can do with data in any approach to modeling. Context-oriented modeling is key to make statistical learning more powerful. Looking back, we statisticians have made significant progresses in the right direction, and we certainly expect more to come. 
% Acknowledgements should go at the end, before appendices and references

\acks{The research is partially supported by the National Science Foundation Awards DMS-1914496 and DMS-2015325.}


\vskip 0.2in
\bibliography{main}

\end{document}